\documentclass[twocolumn, aps, pra, prl, superscriptaddress, preprintnumbers,letterpaper]{revtex4-1}

\usepackage{amsthm}
\usepackage{amsmath,bm}
\usepackage{amssymb,mathrsfs}
\usepackage{braket}
\usepackage{graphicx}
\usepackage{subfigure}
\usepackage{hyperref}
\usepackage{appendix}
\usepackage{tikz}
\usepgflibrary{shapes.geometric}
\usetikzlibrary{arrows}

\renewcommand{\vec}[1]{\boldsymbol{#1}}
\hypersetup{colorlinks=true, citecolor=blue, urlcolor=blue, linkcolor=blue}

\begin{document}
\title{Nonlocal Advantages of Quantum Imaginarity}
\author{Zhi-Wei Wei}
\email{weizhw@cnu.edu.cn}
\affiliation{School of Mathematical Sciences, Capital Normal University, 100048 Beijing, China}
\author{Shao-Ming Fei}
\email{feishm@cnu.edu.cn}
\affiliation{School of Mathematical Sciences, Capital Normal University, 100048 Beijing, China}

\bigskip

\begin{abstract}
Quantum imaginarity and quantum nonlocality capture the essence of quantumness of a physical system from different aspects. We establish a connection between the imaginarity and quantum nonlocality in any two-qubit states. Based on the imaginarity $l_1$-norm and relative entropy, we first derive complementary relations among the quantum imaginarities with respect to any set of mutually unbiased bases for arbitrary qubit states. Based on these complementary relations, we introduce the nonlocal advantage of quantum imaginarity (NAQI).
We show that any two-qubit state with NAQI is quantum steerable, but not the vice-versa. In addition, we investigate the exclusion property of NAQI in three-qubit pure states.
\end{abstract}
\parskip=3pt
\maketitle

\section{introduction}
Quantum mechanics describes the physical reality in terms of complex numbers \cite{stklb1960q,JAraki,caves2001ent,wootters2002p,batle2002ent,batle2003u,PRL102020505,hardy2012l,wootters2012ent,baez2012d,PRA87052106,Wootters2014}. In recent years, the necessity of the imaginary part of quantum mechanics has received great attention \cite{Barnum2020comp,PRA106042207,PRA103L040402,liangxb} and has been rigorously proven by multiple experiments \cite{renou2021q,PRL128040402,PRL128040403}.
It has been shown that the imaginarity can become operationally meaningful in a precise sense \cite{PRL126090401,PRA103032401}.
It has crucial effects in certain discrimination tasks \cite{PRA65050305,wu2023res}, hiding and masking \cite{PRR3033176}, machine learning \cite{PRR5013146}, multiparameter metrology \cite{Carollo2019,Miyazaki2022}, outcome statistics of linear-optical experiments \cite{PRL118153603,PRA98033805}, pseudo-randomness \cite{haug2024pseudorandom}, weak-value theory \cite{HostenKwiat2008,PRL102173601,PRL105010405,PRA85060102,PRA100042116,PRA108L040202} and Kirkwood-Dirac quasiprobability distributions \cite{Wagner2024,PRA108012431,PRA107022408,Budiyono2023}.

Quantum imaginarity is also a valuable resource that cannot be generated under real operations. The resource theory of imaginarity provides insights from a conceptual perspective into the role of the imaginarity played in quantum information theory \cite{PRA95062314,Hickey2018}.
Quantum states whose density matrices contain imaginarity on a pre-fixed reference frame are considered resource states which cannot be freely created by real operations, while the corresponding free states are quantum states that do not contain imaginarity.
In view of resource theory, several imaginarity measures have been proposed such as the geometric imaginarity and the robustness of imaginarity, which have operational significance for state transitions \cite{PRL126090401,PRA103032401}.
This inspired the interest in studying the quantification of localized imaginarity and the related applications in quantum information tasks \cite{xue2021qt,chen2023measures}.
Nevertheless, little attention has been paid to the nonlocal effects of the imaginarity in multipartite systems, which is of importance in remotely controlling and manipulating the imaginarity so as to properly use it as a resource, especially in multipartite scenarios.

In this paper, we study the nonlocality related to the quantum imaginarity in a bipartite scenario by introducing the nonlocal advantage of quantum imaginarity (NAQI). We show that the non-local correlation NAQI is stronger than the steerability, namely, any quantum state with NAQI is steerable, but not the vice-versa. This means that NAQI is indeed a non-local correlation different from the quantum steering. Our results provide an alternate for remotely generating and manipulating the imaginarity of quantum states in noisy environments.

We first adopt two imaginarity measures, the imaginarity $l_1$-norm and relative entropy \cite{xue2021qt,chen2023measures}, and then derive complementary relationships among the quantum imaginarities of a qubit state under mutually unbiased bases \cite{WOOTTERS1989363,bdpdbrv2002new,planat2006survey}.
Based on these complementary relations, we define the NAQI. If certain measurements are made on any subsystem of a bipartite state such that the sum of the average imaginarity of the conditional states of another subsystem violates any quantum imaginarity complementarity relations, we say that this bipartite state has NAQI.
NAQI manifests itself as the steerability of any bipartite quantum states in terms of imaginarity.

In addition, we provide a lower limit on the average imaginarity of any two-qubit systems according to the imaginarity of the reduced states. We further discuss the relationship between NAQI and quantum non-locality, showing that any two-qubit state with NAQI is quantum steerable. By detailed examples, we show that the imaginarity $l_1$-norm can always detect more states with NAQI than the imaginarity relative entropy. In particular, the NAQI is the sufficient for the steerability of quantum states. All quantum states with NAQI are steerable, but not the vice-versa. Finally, for some specific three-qubit pure states, we discuss the exclusion property of NAQI. Numerical testing shows that for the $l_1$-norm of quantum imaginarity, NAQI has exclusion property in these three-qubit pure states.

This paper is structured as follows. In Sec.\ref{sec-1} based on the imaginarity $l_1$-norm and relative entropy, we derive the complementary relations of quantum imaginarity in any set of mutually unbiased bases for any qubit states. In Sec. \ref{sec-2} we introduce the NAQI based on these complementary relations, and provide the corresponding criteria to detect the NAQI of quantum states.
We discuss the exclusion property of NAQI for some specific three-qubit pure states in Sec. \ref{sec-3}, and conclude in the last Sec. \ref{sec-4}.

\section{Complementarity Relations of Quantum Imaginarity}\label{sec-1}

The quantum imaginarity of a state is basis-dependent. Let $\mathcal{B}=\{\ket{a}\}$ ($a=1,2,...,d$) be a pre-selected basis of a $d$-dimensional vector space.
The imaginarity $l_1$-norm $\mathscr{I}^{l_1}_{\mathcal{B}}(\rho)$ of a quantum state $\rho$ is defined by the sum of the absolute values of the imaginary part of all off-diagonal elements of the density matrix $\rho$ \cite{xue2021qt},
\begin{equation}\label{crq1}
\mathscr{I}^{l_1}_{\mathcal{B}}(\rho)=\sum_{a,b}|\mathrm{Im}(\rho_{ab})|,
\end{equation}
where $\rho_{ab}=\braket{a|\rho|b}$ are the entries of $\rho$. The imaginarity $l_1$-norm of $\rho$ is zero if all the off-diagonal elements of the density matrix $\rho$ are real in the basis $\mathcal{B}$. The states with vanishing quantum imaginarity are termed free (real) states in the resource theory of imaginarity. We denote $\mathcal{F}$ the set of all free states.

The imaginarity relative entropy of any quantum state $\rho$ is defined as \cite{chen2023measures}
\begin{equation}\label{crqs1}
\mathscr{I}^{r}_{\mathcal{B}}(\rho)=\min_{\sigma\in\mathcal{F}}S(\rho||\sigma),
\end{equation}
where $\mathcal{F}$ is the set of all free states, $S(\rho||\sigma)=\mathrm{Tr}[\rho\log_{2}{\rho}]-\mathrm{Tr}[\rho\log_{2}{\sigma}]$ is the quantum relative entropy between the quantum states $\rho$ and $\sigma$ \cite{JMPSpDominique}. $\mathscr{I}^{r}_{\mathcal{B}}(\rho)$ can be reexpressed as
\begin{equation}\label{crq2}
\mathscr{I}^{r}_{\mathcal{B}}(\rho)=S(\Delta(\rho))-S(\rho),
\end{equation}
where $S(\rho)=-\mathrm{Tr}[\rho\log_{2}{\rho}]$ is the von Neumann entropy of $\rho$, $\Delta(\rho)=\frac{1}{2}(\rho+\rho^{\mathrm{T}})$ (T stands for the transpose) is a real density matrix given by the real parts of the entries of $\rho$ in the pre-fixed basis $\mathcal{B}$ \cite{xue2021qt}.

Let $M=\{\{\ket{e_i^a}\}^a\}_i$ be a set of bases, where $i$ stands for the $i$th basis and $\ket{e_i^a}$ the $a$th element of the $i$th basis.
A set $M$ of bases is said to be mutual unbiased bases (MUBs) \cite{WOOTTERS1989363,bdpdbrv2002new,planat2006survey} if
$|\braket{e_i^a|e_j^b}|^2=\frac{1}{d}$ for any $i\ne j$ ($i,j=1,2,\cdots,k$) and all $a, b\in\{1,2,\cdots,d\}$, where $k$ represents the number of reference bases in $M$.
Denote $\mathcal{M}_i=\{\ket{e_i^a}\}^a$ the $i$th basis of the set $M$ of MUBs. The set $M$ of MUBs can be expressed as $M=\{\mathcal{M}_i\}_{i=1}^k$.
$M'=\{\mathcal{M'}_i\}_{i=1}^k=\{V\mathcal{M}_iV^\dagger\}_{i=1}^k$ also constitute a set of MUBs for any unitary operator $V$.
Thus, with respect to the sets $M$ and $M'$ of MUBs we have the following relation,
\begin{equation}\label{crq018}
\mathscr{I}_{\mathcal{M}_i}^{\gamma }(\rho)=\mathscr{I}_{V\mathcal{M}_iV^\dagger}^{\gamma }(V\rho V^\dagger),
\end{equation}
where $\gamma$ denotes either $l_1$ or $r$.

For qubit case ($d=2$), there are at most three $(k=3)$ bases in a set $M$ of MUBs, which are given in particular by the eigenbases of the three standard Pauli matrices $\sigma_x, \sigma_y$ and $\sigma_z$. We have the following complementary relations satisfied by the  imaginarity of qubit states $\rho$, see proof in Appendix \ref{a}.

$\mathit{Theorem\ 1}$.
For any qubit state $\rho$, we have
\begin{equation}\label{crq02}
\sum_{i=1}^{3}\mathscr{I}^{\gamma}_{\mathcal{M}_i}(\rho)\leqslant\mathscr{I}_{\gamma}
\end{equation}
with respect to any set $M=\{\mathcal{M}_i\}_{i=1}^3$ of three MUBs, where $\gamma\in\{l_1,r\}$, $\mathscr{I}_{l_1}=\sqrt{5}$ for the imaginarity $l_1$-norm and $\mathscr{I}_{r}\approx2.02685$ associated with the imaginarity relative entropy.

Theorem 1 shows that there are trade-offs among the imaginarities measured by the bases in MUBs for both the imaginarity $l_1$-norm and the imaginarity relative entropy. Namely, if the imaginarity measured in one of the bases in MUBs increases, the sum of the imaginarity measured in the other two bases in MUBs would decrease, and vice-versa.

\section{Nonlocal advantage of quantum imaginarity}\label{sec-2}

As a correlation intermediate between Bell nonlocality \cite{RMP86419} and quantum entanglement \cite{RMP81865}, the quantum steering characterizes the ability of Alice to remotely affect or steer Bob's local state by Alice's arbitrary choice of measurement setting \cite{PR47777,PRL87170405,RMP92015001}, rigorously formulated in terms of local hidden state (LHS) models \cite{PRL98140402,PRA76052116}. The quantum steering has potential applications in many quantum information processing tasks such as one-sided device-independent quantum key distribution \cite{PRA85010301,NC68795,Optica3634}, quantum secret sharing \cite{PRA95010101}, quantum networking tasks \cite{NPhys11167,PRA99012302}, subchannel discrimination \cite{PRL114060404}, quantum teleportation \cite{PRA88062338,PRL115180502} and randomness certification \cite{NJP17113010,QST2015011,PRL120260401}.

Alice prepares an entangled bipartite state $\rho_{AB}$ of systems $A$ and $B$, and sends the subsystem $B$ to Bob. Bob distrusts Alice, but agrees with the fact that the subsystem $B$ is quantum. To convince Bob that the prepared state is indeed entangled and that they share non-local correlations, Alice performs a measurement $x$ and passes the measurement results $a$ to Bob in a classical way. If Alice did not cheat by preparing the subsystem $B$ based on some possible strategies \cite{saunders2010experimental,PRA90050305,PRA92042317}, the conditional state $\rho(a|x)$ obtained by Bob cannot be written by the following LHS model,
\begin{equation}\label{int1}
\rho(a|x)=\sum_{\lambda}p(\lambda)p(a|x,\lambda)\rho_{B}(\lambda)
\end{equation}
for some function $p(a|x,\lambda)$, where $\{p(\lambda),\rho_{B}(\lambda)\}$ is an ensemble of LHS,
$\sum_{\lambda}p(\lambda)=1$ and $\{\rho_{B}(\lambda)\}$ is a given set of quantum states.
Based on Eq. (\ref{int1}), some steering criteria have been derived \cite{saunders2010experimental,PRA90050305,PRA92042317,PRL106130402,PRA87062103,quan2016steering}. Quantitative studies on quantum steering have been given to two-qubit systems and continuous variable systems \cite{PRA93020103,PRL114060403}. Steering scenarios have been demonstrated by a number of experiments with increasing measurement settings \cite{saunders2010experimental} and with loophole-free arrangements \cite{Wittmann2012, smith2012conclusive, PRX2031003, PRX6019902, handchen2012observation, PRA87022104}.

We investigate the effect of quantum imaginarity on quantum nonlocality for any two-qubit systems by using the steering protocol proposed in \cite{PRA95010301,PRA108022410}. Consider a two-qubit state $\rho_{AB}$ shared by Alice and Bob. Alice performs measurements in the bases belonging to a set $\varPi=\{\Pi_i\}_{i=1}^3$ of three arbitrary projective measurements, where $\Pi_i=\{\Pi_i^a\}~(a=1,2)$ stands for the $i$th projective measurement in $\varPi$ and $\Pi_i^a$ the $a$th projective operator of the $i$th projective measurement.
She then communicates her measurement settings and outcomes to Bob, which gives rise to three ensembles of states, $\mathcal{E}_i\equiv\{p(\rho_{B|\Pi_i^a}),\rho_{B|\Pi_i^a}\}$ at Bob's end, where $p(\rho_{B|\Pi_i^a})$ is the probability that Bob obtains the state $\rho_{B|\Pi_i^a}$ after Alice's measurement.
Let $M=\{\mathcal{M}_i\}_{i=1}^3$ be a set of three MUBs. With respect to the $i$th basis $\mathcal{M}_i$ of $M$ the average quantum imaginarity of the ensemble $\mathcal{E}_i$ is given by $\sum_{a}p(\rho_{B|\Pi_i^a})\mathscr{I}_{\mathcal{M}_i}^{\gamma }(\rho_{B|\Pi_i^a})$.
Adding up these average imaginarity over all values of $i$, we have $\sum_{i,a}p(\rho_{B|\Pi_i^a})\mathscr{I}_{\mathcal{M}_i}^{\gamma }(\rho_{B|\Pi_i^a})$.
By maximizing over all the choices of the set $\varPi$ and the set $M$ of MUBs, we define
\begin{equation}\label{crq16-1}
\mathsf{N} _{\gamma }^{\to}(\rho_{AB}):=\max_{M,\varPi}
\sum_{i,a}p(\rho_{B|\Pi_i^a})\mathscr{I}_{\mathcal{M}_i}^{\gamma }(\rho_{B|\Pi_i^a})
\end{equation}
with $\gamma\in\{l_1,r\}$, where the symbol $\to$ stands for that the ensemble generation measurement is carried out on the party $A$, the quantum imaginaritity is calculated on the party $B$, and $\varPi$ is a set of three arbitrary projective measurements, while $M$ is an arbitrary set of three MUBs. For a given quantum state $\rho_{AB}$, the calculation of $\mathsf{N}_{\gamma }^{\to}(\rho_{AB})$ defined in Eq.(\ref{crq16-1}) is illustrated in Appendix \ref{ac}.

[{\it{Definition}}]
For any two qubit state $\rho_{AB}$, if the average imaginarity $\mathsf{N} _{\gamma }^{\to}(\rho_{AB})$ violates the quantum imaginary complementarity relations (\ref{crq02}), i.e.,
\begin{equation}\label{crq16}
\mathsf{N} _{\gamma }^{\to}(\rho_{AB})>\mathscr{I}_{\gamma}
\end{equation}
for any $\gamma\in\{l_1,r\}$, we say that the state $\rho_{AB}$ has non-local advantage of quantum imaginarity (NAQI).

By definition we see that NAQI is a kind of measurement induced non-local correlation.
We have the following conclusions, see proof in Appendix \ref{b}.

$\mathit{Theorem\ 2}$.
For any two-qubit state $\rho_{AB}$, $\mathsf{N}_{\gamma }^{\to}(\rho_{AB})$ is invariant under local unitary transformations, and lower bounded by the sum of the imaginarities of the reduced state $\rho_B=\mathrm{Tr}_A[\rho_{AB}]$ with respect to any set $M=\{\mathcal{M}_i\}_{i=1}^3$ of three mutually unbiased bases,
\begin{equation}\label{thm2}
\mathsf{N}_{\gamma }^{\to}(\rho_{AB})\geqslant\max_{M}\sum_{i=1}^3\mathscr{I}_{\mathcal{M}_i}^{\gamma }(\rho_B).
\end{equation}

The NAQI presents a sufficient criterion on the steerability of a state from Alice to Bob.

$\mathit{Theorem\ 3}$.
Any two-qubit state $\rho_{AB}$ with NAQI, namely, $\mathsf{N} _{\gamma }^{\to}(\rho_{AB})>\mathscr{I}_{\gamma}$ for any $\gamma\in\{l_1,r\}$, is quantum steerable.

$\mathit{Proof}$.
Let $\rho_{AB}$ be an arbitrary two-qubit state with NAQI. Assume that $\rho_{AB}$ is not a quantum steerable state, i.e., it admits an LHS model defined in (\ref{int1}).
For any imaginarity measure $\mathscr{I}_{\mathcal{B}}^{\gamma }$, $\gamma\in\{l_1,r\}$, where $\mathcal{B}$ is a pre-fixed reference basis.
By replacing $\mathcal{B}$ with different reference bases $\mathcal{M}_i~(i=1,2,3)$ in $M$, we have
\begin{align}\label{crq15}
&\mathsf{N}_{\gamma }^{\to}(\rho_{AB})\nonumber\\
&=\max_{M,\varPi}\sum_{i,a}p(\rho_{B|\Pi_i^a})\mathscr{I}_{\mathcal{M}_i}^{\gamma }(\rho_{B|\Pi_i^a})\nonumber\\
&=\max_{M,\varPi}\sum_{i,a}p(\rho_{B|\Pi_i^a})\mathscr{I}_{\mathcal{M}_i}^{\gamma }(\frac{\sum_{\lambda}p(\lambda)p(a|\Pi_i^a,\lambda)\rho_{B}(\lambda)}{p(\rho_{B|\Pi_i^a})})\nonumber\\
&\leqslant\max_{M,\varPi}\sum_{i, a,\lambda}p(\lambda)p(a|\Pi_i^a,\lambda)\mathscr{I}_{\mathcal{M}_i}^{\gamma }(\rho_{B}(\lambda))\nonumber\\
&=\max_{M}\sum_{i, \lambda}p(\lambda)\mathscr{I}_{\mathcal{M}_i}^{\gamma }(\rho_{B}(\lambda))\nonumber\\
&\leqslant\sum_{\lambda}p(\lambda)\mathscr{I}_{\gamma}=\mathscr{I}_{\gamma},
\end{align}
where $\mathscr{I}_{l_1}$ and $\mathscr{I}_{r}$ are given in Theorem 1,
the first and the second inequalities are obtained, respectively, by utilizing the convexity of the quantum imaginarity and the imaginarity complementarity relation for qubit systems (\ref{crq02}). This contradicts with the assumption (\ref{crq16}).
Therefore, any two-qubit state $\rho_{AB}$ with NAQI is quantum steerable.
$\hfill\qedsymbol$

Theorem 3 shows that any two-qubit state with NAQI is steerable. The NAQI captures a kind of quantum correlation which is stronger than steerability. It provides an alternate for remotely generating and manipulating the imaginarity of quantum states in noisy environments. Similar to the nonlocal advantage of quantum coherence (NAQC) \cite{PRA95010301,PRA108022410,PRA106012433,PRA101032305,liu2017q}, NAQI contains more information than quantum imaginarity itself in the sense that it characterizes the steerability of a bipartite states.
Any state with quantum imaginarity with respect to a fixed basis must be coherent, but not vice-versa \cite{PRL113140401,PRL116150502,PRL116120404,RMPCoh}. For certain quantum states, the NAQC may be not detected via the coherence complementarity relation \cite{PRA95010301}, but the NAQI may be detected via the imaginary complementarity relation, since the upper limit of the quantum imaginary complementarity relation is significantly smaller than the upper limit of the quantum coherence complementarity relation.
Namely, there are quantum states with NAQI but not with NAQC. Therefore, the NAQI is stronger than steerability but weaker than NAQC in general.
We give two specific examples to illustrate the NAQI defined by (\ref{crq16}).

{\sf [Example 1]} Consider the classical probabilistic mixture of two Bell states,
\begin{equation}\label{exam1}
\rho_{AB}=p|\phi^+\rangle\langle\phi^+|+(1-p)|\psi^+\rangle\langle\psi^+|,
\end{equation}
where $\ket{\phi^+}=\frac{1}{\sqrt{2}}(\ket{00}+\ket{11})$ and $\ket{\psi^+}=\frac{1}{\sqrt{2}}(\ket{10}+\ket{01})$, $p\in[0,1]$.
$\rho_{AB}$ is Bell nonlocal and with NAQC for any $p\ne0.5$ \cite{PRA108022410}.

\begin{figure}[t]
  \centering
  \includegraphics[width=8cm]{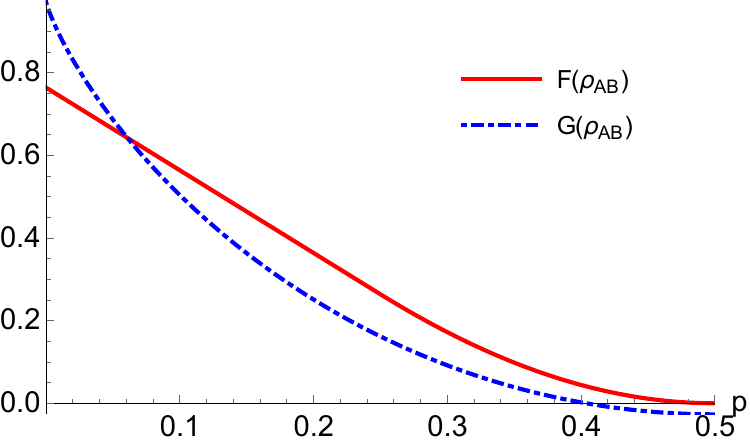}\\
  \caption{NAQI for probabilistic mixtures of two Bell states. The solid (red) line denotes the functional $\mathsf{F}(\rho_{AB})$ of the  imaginarity $l_1$-norm. The dotdashed (blue) line denotes the functional $\mathsf{G}(\rho_{AB})$ of the imaginarity relative entropy. All quantities used are dimensionless.}
  \label{f1}
\end{figure}

By adopting the $l_1$-norm and the relative entropy as measures of imaginarity, i.e., $\forall~\gamma\in\{l_1,r\}$, we calculate the functions $\mathsf{F}(\rho_{AB})=\mathsf{N}_{l_1}^{\to}(\rho_{AB})-\mathscr{I}_{l_1}$ and $\mathsf{G}(\rho_{AB})=\mathsf{N}_{r}^{\to}(\rho_{AB})-\mathscr{I}_{r}$ respectively, which are symmetric about $p=0.5$. According to the definition of NAQI by (\ref{crq16}), either $\mathsf{F}(\rho_{AB})>0$ or $\mathsf{G}(\rho_{AB})>0$ indicate that $\rho_{AB}$ has NAQI. In Fig. \ref{f1}, it is seen that the quantum state $\rho_{AB}$ given by (\ref{exam1}) has NAQI for $p\ne0.5$ in terms of the imaginarity $l_1$-norm, and has NAQI for $p\lesssim0.403$ or $p\gtrsim0.597$ based on the imaginarity relative entropy.

{\sf [Example 2]} Let us consider the two-qubit Werner state,
\begin{equation}\label{exam2}
\rho_{W}=p|\phi^+\rangle\langle\phi^+|+\frac{1-p}{4}\mathbb{I}\otimes\mathbb{I},
\end{equation}
where $p\in[0,1]$ and $\mathbb{I}$ denotes the identity operator on $\mathbb{C}^2$.
State $\rho_{W}$ is entangled for $p>\frac{1}{3}$ \cite{PRL771413,HORODECKI19961}, steerable for $p >\frac{1}{2}$ \cite{PRL98140402}, Bell nonlocal for $p>\frac{1}{\sqrt{2}}$, and NAQC for $p\gtrsim0.815$ \cite{PRA108022410}.

\begin{figure}[t]
  \centering
  \includegraphics[width=8cm]{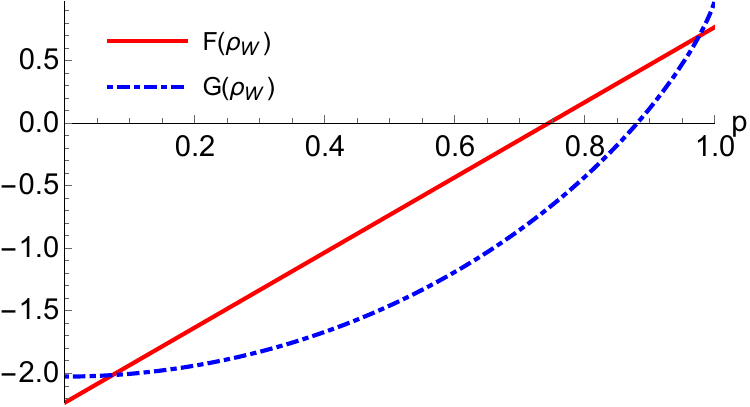}\\
  \caption{NAQI for two-qubit Werner states. The solid (red) line is for the function $\mathsf{F}(\rho_{W})$ of the imaginarity $l_1$-norm. The dotdashed (blue) line is for the function $\mathsf{G}(\rho_{W})$ of the imaginarity relative entropy. All quantities used are dimensionless.}
  \label{f2}
\end{figure}

With respect to the imaginarity $l_1$-norm and relative entropy, we calculate $\mathsf{F}(\rho_{W})=\mathsf{N}_{l_1}^{\to}(\rho_{W})-\mathscr{I}_{l_1}$ and $\mathsf{G}(\rho_{W})=\mathsf{N}_{r}^{\to}(\rho_{W})-\mathscr{I}_{r}$ respectively.
Similarly, either $\mathsf{F}(\rho_{W})>0$ or $\mathsf{G}(\rho_{W})>0$ indicate that the state $\rho_{W}$ has NAQI.
Fig. \ref{f2} shows that $\rho_{W}$ has NAQI for $p\gtrsim0.7454$ based on the imaginarity $l_1$-norm, and for $p\gtrsim0.8816$ based on the imaginarity relative entropy.

Interestingly, we observe in Fig. \ref{f2} that there exist steerable and Bell nonlocal states which do not exhibit NAQI when the imaginarity $l_1$-norm or the imaginarity relative entropy are taken into account. Formula (\ref{crq16}) can also be regarded as two criteria of NAQI, in which the imaginary $l_1$-norm can always detect more quantum states with NAQI than the imaginary relative entropy.
This implies that the two NAQI criteria constructed from the two local imaginarity measures capture different sets of NAQI states.
In particular, as for two-qubit Werner states, we find that the set of states with NAQI under the NAQI criterion of the imaginarity $l_1$-norm is larger than the set of states with NAQC \cite{PRA108022410}. In addition, it can be seen from these two specific examples that NAQI may be stronger than Bell nonlocality.

\section{Exclusion of NAQI}\label{sec-3}
We now study the exclusion of the NAQI characterized by $\mathsf{N}_{\gamma }^{\to}(\rho_{AB})$ in (\ref{crq16-1}) for some specific three-qubit pure states. A three-qubit pure state $\ket{\Psi_{ABC}}$ can be represented on the computational basis as follows,
\begin{align}\label{sec3-0}
\ket{\Psi_{ABC}}=&\lambda_0\ket{000}+\lambda_1e^{i\phi}\ket{100}+\lambda_2\ket{101}\nonumber\\
&+\lambda_3\ket{110}+\lambda_4\ket{111},
\end{align}
where $\lambda_i\geqslant0$, $\sum_{i=0}^{4}\lambda_i^2=1$ and $\phi\in[0,\pi]$ \cite{PRL851560,AAc2001}. $\ket{\Psi_{ABC}}$ reduces to the W-class states \cite{PRA62062314,PRL87040401} when $\lambda_4=0$ and $\phi=0$.

Let $\rho_{AB}$, $\rho_{BC}$ and $\rho_{CA}$ denote the reduced states of $\rho_{ABC}=|\Psi_{ABC}\rangle\langle\Psi_{ABC}|$ obtained by tracing over  parties $C$, $A$ and $B$, respectively.
$\mathsf{N}_{\gamma }^{\to}(\rho_{AB})$ involves quantum measurements on one of the subsystems and the imaginarity of the reduced states on another subsystem.
Hence, we assume that the projection measurements of the reduced states $\rho_{AB}$, $\rho_{BC}$ and $\rho_{CA}$ are performed on the subsystems $A$, $B$ and $C$, respectively, and the correspondingly the imaginarity are calculated on the subsystems $B$, $C$ and $A$ respectively. We only examine the exclusion of the NAQI defined in (\ref{crq16}) based on the imaginarity $l_1$-norm for some specific three-qubit pure states $\ket{\Psi_{ABC}}$ given in (\ref{sec3-0}).

\begin{figure}[t]
  \centering
  \includegraphics[width=8cm]{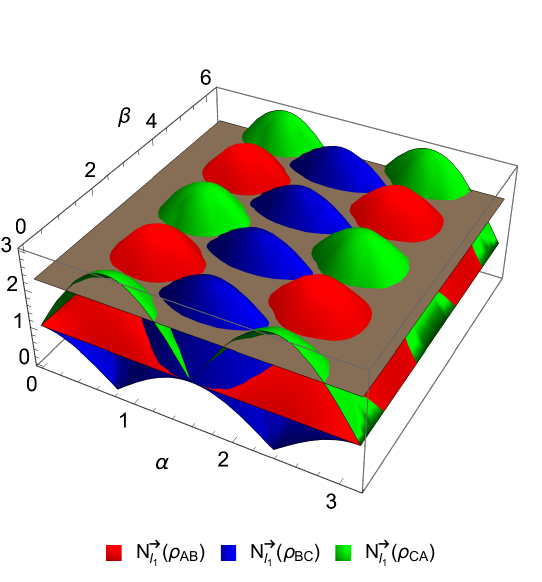}\\
  \caption{NAQI tests based on the imaginarity $l_1$-norm. The red (blue) surface stands for  $\mathsf{N}_{l_1}^{\to}(\rho_{AB})$ ($\mathsf{N}_{l_1}^{\to}(\rho_{BC})$),  the green surface is for the $\mathsf{N}_{l_1}^{\to}(\rho_{CA})$. The gray surface indicates the critical plane of NAQI. Whenever one of the three pairs of qubits has NAQI, the other two pairs of qubits do not have NAQI. All quantities used are dimensionless.}
  \label{f3}
\end{figure}

\begin{figure}[t]
  \centering
  \includegraphics[width=8cm]{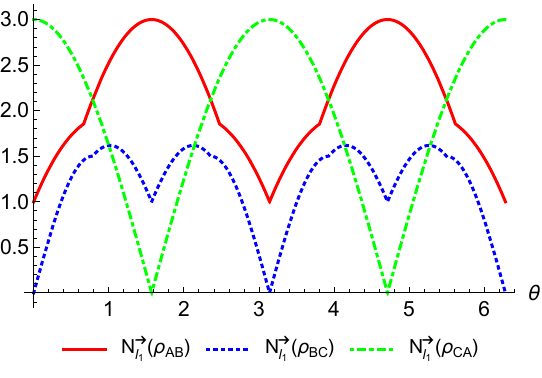}\\
  \caption{NAQI tests based on the imaginarity $l_1$-norm. The solid (red) line denotes $\mathsf{N}_{l_1}^{\to}(\rho_{AB})$,  the dotted (blue) line is for $\mathsf{N}_{l_1}^{\to}(\rho_{BC})$,  and the dot dashed (green) stands for $\mathsf{N}_{l_1}^{\to}(\rho_{CA})$. All quantities used are dimensionless.}
  \label{f4}
\end{figure}

Let us set $\lambda_0=\cos\alpha, \lambda_2=\sin\alpha\cos\beta, \lambda_3=\sin\alpha\sin\beta$ and $\lambda_1=\lambda_4=\phi=0$  in (\ref{sec3-0}), where $\alpha\in[0,\pi]$ and $\beta\in[0,2\pi]$. Fig. \ref{f3} shows the exclusion property among the NAQIs on pairwise subsystems.
By choosing $\lambda_0=\frac{\sqrt{2}}{2}$, $\lambda_2=\frac{\sqrt{2}}{2}\cos\theta$ and $\lambda_3=\frac{\sqrt{2}}{2}\sin\theta$, we obtain the plane graph (Fig. \ref{f4}). From Fig. \ref{f4} one sees that when one pair of qubits has NAQI, the other two pairs of qubits can no longer have NAQI.
It is impossible that all pairs of qubits have NAQI simultaneously. Such phenomena is similar to the trade-offs among the violations Bell inequalities in three-qubit states \cite{PRA92062339}.

\section{Conclusion}\label{sec-4}
We have derived the quantum imaginarity complementary relations (\ref{crq02}) satisfied by the imaginarity with respect to references in any set of mutually unbiased bases for any qubit states, based on the imaginarity $l_1$-norm and relative entropy.
By performing local measurements on $A$ part of any two-qubit state $\rho_{AB}$, it is shown that the average imaginarity of the conditional states of $B$ part, summing over all the mutually unbiased bases, can exceed a threshold that cannot be exceeded by any qubit state. Based on these complementary relations, we have introduced a kind of NAQI given by (\ref{crq16}), which is invariant under local unitary operations.
In addition, we have provided a lower limit on the function $\mathsf{N}_{\gamma}^{\to}(\rho_{AB})$ of any two-qubit systems according to imaginarity of the reduced states.

We have further discussed the relationship between the NAQI and other quantum non-locality, and shown that any two-qubit state with NAQI is quantum steerable. Therefore, NAQI captures a kind of quantum correlation which is stronger than quantum steerability. All quantum states with NAQI are steerable, but not the vice-versa. In particular, the NAQI given by (\ref {crq16}) can also be regarded as a steering witness based on local quantum imaginarity. By detailed examples, it has been shown that the imaginarity $l_1$-norm can always detect more quantum states with NAQI than the imaginarity relative entropy. Finally, for some specific three-qubit pure states, we have discussed the exclusion principle of NAQI. Numerical testing shows that NAQI based on the imaginarity $l_1$-norm has exclusion property in these specific three-qubit pure states.

Our results only reveal the important connection between the imaginarity and quantum nonlocal correlation for any two-qubit states. The quantum imaginarity complementarity relations for general high-dimensional bipartite states in any set of mutually unbiased bases need to be explored.
Here it should be noted that, as a kind of quantum nonlocal correlation induced by local imaginarity, the NAQI is independent of the detailed imaginarity measures. Nevertheless, our criteria of detecting NAQI, constructed from detailed local imaginarity measures, are measure dependent, as different measures of imaginarity capture different sets of NAQI states. In addition, we have shown that NAQI is stronger than quantum steering, and naturally stronger than entanglement.
Although the examples show that NAQI may be even stronger than the Bell nonlocality, the hierarchy of NAQI and Bell nonlocality still needs further study.

Our results give a deeper understanding of the interrelation between quantum imaginarity and quantum correlations in composite systems, and provide some clues in revealing the interplay between the unavoidable deimaginarity of a system and efficiency of the active quantum operations on protecting imaginarity.
This arises naturally in two-qubit systems in which both parties work together to generate the largest possible local imaginarity on one of the subsystems. Consider that Alice and Bob share a two-qubit state. The imaginarity is then a resource in Bob's system. Alice may perform arbitrary local projective measurements on her system and communicate classically with Bob, so that the conditional states of Bob's system achieve an NAQI.
From a practical point of view, our results also present an alternate for remotely generating and manipulating imaginarity in noisy environments.
The NAQI may contain more comprehensive information than that of imaginarity. Therefore, it has the potential to distinguish the subtle nature of a quantum system and more reliably reflects the quantum critical behaviors even when the imaginarity measures fail to.

\bigskip
\section*{Acknowledgments}
This work is supported by the National Natural Science Foundation of China under Grant Nos. 12075159 and 12171044; the specific research fund of the Innovation Platform for Academicians of Hainan Province.

\begin{appendix}
\section{Proof of Theorem 1}\label{a}
Any qubit state $\rho=\frac{1}{2}(\mathbb{I}+\vec{n}\cdot\vec{\sigma})$ in the reference basis $z$ given by the eigenvectors of the Pauli matrix $\sigma_z$ can be written as
\begin{equation}\label{crq3}
\rho=\frac{1}{2}\begin{pmatrix}
 1+n_z & n_x-in_y\\
 n_x+in_y&1-n_z
\end{pmatrix},
\end{equation}
where $\vec{n}=(n_x, n_y, n_z)\in\mathbb{R}^3$ with $|\vec{n}|=\sqrt{n_x^2+n_y^2+n_z^2}\leqslant1$, $\vec{\sigma}\equiv(\sigma_x,\sigma_y,\sigma_z)$ is the Pauli matrix vector and $\mathbb{I}$ is the $2\times2$ identity matrix.

Let us first consider the imaginarity $l_1$-norm defined in (\ref{crq1}) for any qubit state $\rho$. It can be seen from the formula (\ref{crq3}) that the imaginarity $l_1$-norm of $\rho$ in the reference basis $z$ is given by
\begin{equation}\label{crq4}
\mathscr{I}^{l_1}_{z}(\rho)=|n_y|.
\end{equation}
Denote the reference bases given by the eigenvectors of the Pauli matrices $\sigma_x$ and $\sigma_y$ by $x$ and $y$, respectively. By simple calculation, the imaginarity $l_1$-norm of $\rho$ is given by $\mathscr{I}^{l_1}_{x}(\rho)=|n_y|$ and $\mathscr{I}^{l_1}_{y}(\rho)=|n_x|$ in the reference bases $x$ and $y$, respectively.

Subsequently, we consider the upper limit on $\mathscr{I}^{l_1}_{x}(\rho)+\mathscr{I}^{l_1}_{y}(\rho)+\mathscr{I}^{l_1}_{z}(\rho)=|n_x|+2|n_y|$ for any qubit state $\rho$.
Using the Cauchy-Schwarz inequality, we have
\begin{equation}\label{crq5}
\sum_{i=x,y,z}\mathscr{I}^{l_1}_{i}(\rho)
\leqslant\sqrt{5}\sqrt{n_x^2+n_y^2}\leqslant\sqrt{5}=\mathscr{I}_{l_1},
\end{equation}
where the equality applies to the pure state $n_x=\frac{1}{\sqrt{5}},n_y=\frac{2}{\sqrt{5}}$ and $n_z=0$. Hence, $|n_x|+2|n_y|$ is no greater than $\sqrt{5}$, and the inequality (\ref{crq5}) is called the complementarity relation of the imaginarity $l_1$-norm.

Next, we consider the relative entropy of imaginarity defined by (\ref{crq2}) for any qubit state $\rho$. From the matrix form (\ref{crq3}) of $\rho$ in the reference basis $z$, the corresponding real density matrix $\Delta(\rho)$ can be written as
\begin{equation}\label{crq6}
\Delta_z(\rho)=\frac{1}{2}\begin{pmatrix}
 1+n_z & n_x\\
 n_x&1-n_z
\end{pmatrix}.
\end{equation}
Therefore, the imaginarity relative entropy of $\rho$ in the reference basis $z$ is given by
\begin{equation}\label{crq7}
\mathscr{I}^{r}_{z}(\rho)=H(\frac{1+\sqrt{n_x^2+n_z^2}}{2})-H(\frac{1+|\vec{n}|}{2}),
\end{equation}
where $H(x)=-x\log_{2}{x}-(1-x)\log_{2}{(1-x)}$.

Similarly, with respect to the basis $x$ we have
\begin{equation}\label{crq8}
\Delta_x(\rho)=\frac{1}{2}\begin{pmatrix}
 1+n_x & -n_z\\
 -n_z&1-n_x
\end{pmatrix},
\end{equation}
\begin{equation}\label{crq9}
\mathscr{I}^{r}_{x}(\rho)=H(\frac{1+\sqrt{n_x^2+n_z^2}}{2})-H(\frac{1+|\vec{n}|}{2}),
\end{equation}
and with respect to the basis $y$ we have
\begin{equation}\label{crq10}
\Delta_y(\rho)=\frac{1}{2}\begin{pmatrix}
 1+n_y & -n_z\\
 -n_z&1-n_y
\end{pmatrix},
\end{equation}
\begin{equation}\label{crq11}
\mathscr{I}^{r}_{y}(\rho)=H(\frac{1+\sqrt{n_y^2+n_z^2}}{2})-H(\frac{1+|\vec{n}|}{2}).
\end{equation}

From (\ref{crq7}), (\ref{crq9}) and (\ref{crq11}), the sum of relative entropy of imaginarity for any qubit state $\rho$ in reference bases $x,y,z$ is bounded by
\begin{align}\label{crq12}
\sum_{i=x, y, z}\mathscr{I}^{r}_{i}(\rho)=&2H(\frac{1+\sqrt{n_x^2+n_z^2}}{2})
+H(\frac{1+\sqrt{n_y^2+n_z^2}}{2})\nonumber\\
&-3H(\frac{1+|\vec{n}|}{2})\nonumber\\
\leqslant&2H(\frac{1+\sqrt{n_x^2+n_z^2}}{2})+H(\frac{1+\sqrt{n_y^2+n_z^2}}{2})\nonumber\\
\leqslant&\mathscr{I}_{r},
\end{align}
where the first inequality holds an equal sign for any pure state. For the second inequality, by numerical calculations it is easy to know that the maximum $\mathscr{I}_{r}\approx2.0269$ attains at the pure state given by $n_x\approx0.27249, n_y\approx0.96216$ and $n_z=0$.
Combining (\ref{crq5}) and (\ref{crq12}), we obtain
\begin{equation}\label{acrq2}
\sum_{i=x,y,z}\mathscr{I}^{\gamma}_{i}(\rho)\leqslant\mathscr{I}_{\gamma}
\end{equation}
for any qubit state $\rho$, where $\gamma\in\{l_1,r\}$, $\mathscr{I}_{l_1}=\sqrt{5}$ and $\mathscr{I}_{r}\approx2.02685$.

According to relation (\ref{crq018}), for any qubit state $\rho$, we get
\begin{equation}\label{acrq3}
\sum_{i=1}^{3}\mathscr{I}^{\gamma}_{\mathcal{M}_i}(\rho)\leqslant\mathscr{I}_{\gamma}
\end{equation}
with respect to any set $M=\{\mathcal{M}_i\}_{i=1}^3$ of three MUBs, where $\gamma\in\{l_1,r\}$, $\mathscr{I}_{l_1}=\sqrt{5}$ for the imaginarity $l_1$-norm and $\mathscr{I}_{r}\approx2.02685$ associated with the imaginarity relative entropy.

\section{Calculation of $\mathsf{N} _{\gamma }^{\to}(\rho_{AB})$}\label{ac}
For a given quantum state $\rho_{AB}$, according to the definition of $\mathsf{N} _{\gamma }^{\to}(\rho_{AB})$ given in Eq.(\ref{crq16-1}), the maximum goes over all different mutually unbiased basis sets $M$ and all choices of three arbitrary projection measurement sets $\varPi$.
By parametric representation, the set $M=\{\mathcal{M}_i\}_{i=1}^3$ with $\mathcal{M}_i=\{\mathcal{M}_i^{\pm}\}$ can be represented as
\begin{align}
\{\ket{\mathcal{M}_1^{\pm}}\}=&\left\{\cos\frac{\theta_1}{2}\ket{0}+e^{i\phi_1}\sin\frac{\theta_1}{2}\ket{1},\right.\nonumber\\
&\left.\sin\frac{\theta_1}{2}\ket{0}-e^{i\phi_1}\cos\frac{\theta_1}{2}\ket{1}\right\},\\
&\{\ket{\mathcal{M}_2^{\pm}}\}=\left\{\frac{\ket{\mathcal{M}_1^{+}}\pm\ket{\mathcal{M}_1^{-}}}{\sqrt{2}}\right\},\\
&\{\ket{\mathcal{M}_3^{\pm}}\}=\left\{\frac{\ket{\mathcal{M}_1^{+}}\pm i\ket{\mathcal{M}_1^{-}}}{\sqrt{2}}\right\},
\end{align}
where $\ket{0}$ and $\ket{1}$ are the eigenvectors of the Pauli matrix $\sigma_z$, $\theta_1$ and $\phi_1$ are the azimuthal and polar angles in spherical polar coordinates. The set $\varPi=\{\Pi_i\}_{i=1}^3$ with $\Pi_i=\{\Pi_i^a\}~(a=1,2)$ can be represented as
\begin{align}
\{\Pi_i^a\}=&\left\{\cos\frac{\theta_{i+1}}{2}\ket{0}+e^{i\phi_{i+1}}\sin\frac{\theta_{i+1}}{2}\ket{1},\right.\nonumber\\
&\left.\sin\frac{\theta_{i+1}}{2}\ket{0}-e^{i\phi_{i+1}}\cos\frac{\theta_{i+1}}{2}\ket{1}\right\}
\end{align}
for each $i=1,2,3$, where $\{\theta_2$, $\phi_2\}, \{\theta_3$, $\phi_3\}$ and $\{\theta_4$, $\phi_4\}$ are three independent sets of azimuthal and polar angles in spherical polar coordinates. Since the sets $M$ and $\varPi$ are independent of each other, $\mathsf{N} _{\gamma }^{\to}(\rho_{AB})$ is the maximum of the function defined in (\ref{crq16-1}) over the eight independent variables.

\section{Proof of Theorem 2}\label{b}
We first prove that the function defined by formula (\ref{crq16-1}) is invariant under local unitary transformation.
Consider the transformation $\rho'_{AB}=(U\otimes V)\rho_{AB}(U^\dagger\otimes V^\dagger)$, where $U$ and $V$ are arbitrary unitary operators.
We only need to prove that $\mathsf{N}_{\gamma }^{\to}(\rho_{AB})=\mathsf{N}_{\gamma }^{\to}(\rho'_{AB})$.

Suppose  $\varPi=\{\Pi_i\}_{i=1}^3$, $\Pi_i=\{\Pi_i^a\}~(a=1,2)$, is a set of three arbitrary projective measurements and $M=\{\mathcal{M}_i\}_{i=1}^3$ an arbitrary set of three MUBs that maximize the formula (\ref{crq16-1}) for the state $\rho_{AB}$.
Let $\rho_{B|\Pi_i^a}$ be the conditional state with probability $p(\rho_{B|\Pi_i^a})$ at Bob's side when Alice performs $\Pi_i^a$ measurement.

For state $\rho'_{AB}$, consider the set $\varPi'=\{\Pi'_i\}_{i=1}^3$ with $\Pi'_i=\{\Pi'^a_i\}=\{U\Pi_i^aU^\dagger\}~(a=1,2,i=1,2,3)$ of the three projective measurements performed by Alice and the set $M'=\{\mathcal{M}'_i\}_{i=1}^3=\{V\mathcal{M}_iV^\dagger\}_{i=1}^3$ of the three MUBs used by Bob to calculate $\sum_{i, a}p(\rho'_{B|\Pi'^a_i})\mathscr{I}_{\mathcal{M}'_i}^{\gamma }(\rho'_{B|\Pi'^a_i})$.
For every $i$, the reduced states at Bob's end, created due to measurements $\Pi'_i$ by Alice on $\rho'_{AB}$, are $\rho'_{B|\Pi'^a_i}=V\rho_{B|\Pi_i^a}V^\dagger$ $(a=1,2)$ and the corresponding occurrence probability are $p(\rho'_{B|\Pi'^a_i})=p(\rho_{B|\Pi_i^a})$.
Notice the relationship in Eq.(\ref{crq018}), it is obvious that
\begin{equation}\label{crq19}
\sum_{i, a}p(\rho'_{B|\Pi'^a_i})\mathscr{I}_{\mathcal{M}'_i}^{\gamma}(\rho'_{B|\Pi'^a_i})=\mathsf{N}_{\gamma}^{\to}(\rho_{AB}).
\end{equation}
Hence, according to the definition (\ref{crq16-1}), we have
\begin{equation}\label{crq20}
\mathsf{N}_{\gamma }^{\to}(\rho'_{AB})\geqslant\mathsf{N}_{\gamma}^{\to}(\rho_{AB}).
\end{equation}

On the other hand, consider $\rho_{AB}=(U^\dagger\otimes V^\dagger)\rho'_{AB}(U\otimes V)$, where $U^\dagger$ and $V^\dagger$ are also unitary. Similar treatment as above leads to
\begin{equation}\label{crq21}
\mathsf{N}_{\gamma }^{\to}(\rho_{AB})\geqslant\mathsf{N}_{\gamma }^{\to}(\rho'_{AB}).
\end{equation}
From (\ref{crq20}) and (\ref{crq21}), we have
\begin{equation}\label{crq23}
\mathsf{N}_{\gamma }^{\to}(\rho_{AB})=\mathsf{N}_{\gamma }^{\to}(\rho'_{AB}).
\end{equation}
Hence, we have proven invariance of $\mathsf{N}_{\gamma }^{\to}(\rho_{AB})$ under local unitaries.

Finally, let $M=\{\mathcal{M}_i\}_{i=1}^3$ be any set of three MUBs and $\varPi=\{\Pi_i\}_{i=1}^3$ with $\Pi_i=\{\Pi_i^a\}~(a=1,2,i=1,2,3)$ be any set of three arbitrary projective measurements.
According to the definition (\ref{crq16-1}), we have
\begin{align}\label{crq24}
\mathsf{N}_{\gamma }^{\to}(\rho_{AB})&=\max_{M,\varPi}\sum_{i,a}p(\rho_{B|\Pi_i^a})\mathscr{I}_{\mathcal{M}_i}^{\gamma }(\rho_{B|\Pi_i^a})\nonumber\\
&\geqslant\max_{M}\sum_{i}\mathscr{I}_{\mathcal{M}_i}^{\gamma }(\rho_B),
\end{align}
where the inequality is due to the convexity of quantum imaginarity and the fact that any measurement on any subsystem of a system cannot disturb the average state of the other subsystems, i.e., $\sum_{a}p(\rho_{B|\Pi_i^a})\rho_{B|\Pi_i^a}=\mathrm{Tr}_A[\rho_{AB}]=\rho_B$.

\end{appendix}


\end{document}